\documentclass{llncs}
\usepackage{soul}
\usepackage{graphicx}
\usepackage{float}
\usepackage{amsmath}
\usepackage[normalem]{ulem}
\usepackage{fancyvrb}
\usepackage{cite}
\usepackage{spverbatim}
\usepackage{url}

\begin{document}

\title{Mining State-Based Models from Proof Corpora\thanks{The final publication is available at http://link.springer.com}}
\author{Thomas Gransden \and Neil Walkinshaw \and Rajeev Raman}
\institute{Department of Computer Science, University of Leicester, Leicester, UK\\
\email{tg75@student.le.ac.uk}, \email{n.walkinshaw@mcs.le.ac.uk}, \email{r.raman@le.ac.uk}}
\maketitle

\begin{abstract}
Interactive theorem provers have been used extensively to reason about various software/hardware systems and mathematical theorems. The key challenge when using an interactive prover is finding a suitable sequence of proof steps that will lead to a successful proof requires a significant amount of human intervention. This paper presents an automated technique that takes as input examples of successful proofs and infers an Extended Finite State Machine as output. This can in turn be used to generate proofs of new conjectures. Our preliminary experiments show that the inferred models are generally accurate (contain few false-positive sequences) and that representing existing proofs in such a way can be very useful when guiding new ones.
\end{abstract}

\section{Introduction}
\label{sec:introduction}
Interactive theorem provers (ITPs) provide a semi-automatic environment in which a user can reason about the correctness of hardware and software systems and verify the proofs of significant mathematical theorems. Given a desired property expressed in a formal logic, provers such as Coq \cite{Coq:manual} and Isabelle \cite{Isabelle02} provide a framework by which to construct higher-order logic proofs in a stepwise manner, drawing upon libraries of existing proven theorems. In the context of computer mathematics, ITPs have successfully been used in the verification of the Four Color Theorem \cite{Gonthier07a}, the Kepler Conjecture \cite{Hales05} and the Feit-Thompson Theorem \cite{Gonthier13}.

ITPs rely on the ability of an expert to choose suitable proof steps to apply. Clearly this requires not only the selection of the correct proof steps, but also knowledge about how to sequence these proof steps in order to arrive at a successful proof. To complicate matters further, the user must select suitable parameters for these proof steps. In a significant development, the overall proof effort can contain tens of thousands of lines. For example, Gonthier's machine checked proof of the Feit-Thompson theorem amounted to 170,000 lines of code. This shows that a lot of human effort was needed to complete the proof.

Over the past decade, several semi-automatic tools have been developed to simplify the verification process \cite{ML4PG13,Duncan07,Dixon03,Grov13}. These tools adopt data mining and heuristic search strategies to identify proof patterns and to conjecture new proofs. One outstanding challenge, recently highlighted by Grov \emph{et al.} \cite{Grov12}, is the need to identify \emph{proof strategies}. There is a desire not only to recognize common syntactic patterns (as achieved by current techniques), but to take this one step further and to capture the rules that govern the possible ordering of the proof steps required to yield a successful proof. This is what motivates the work in this paper.

Accordingly, we present a technique to derive sequential models (in the form of Extended Finite State Machines (EFSM) \cite{Walkinshaw13}) from existing corpora of proofs. These models can be interpreted as an instance of the proof strategies referred to by Grov \emph{et al}. The models capture the reasoning patterns that bind groups of proofs together, and in doing so capture the possible sequences of proof steps that have led to successful proofs. Corpora that contain tens or hundreds of proofs can be collapsed into (relatively) compact, graphical models. We show how these models can be used to the benefit of interactive theorem prover users. The specific contributions of this paper are:
\begin{itemize}
\item{A technique to automatically derive EFSM models from libraries of interactive proofs (Section \ref{sec:efsm}).}
\item{An evaluation that indicates that the models are broadly precise and can be used as an aid to yield proofs of new propositions, and to shorten existing ones (Section \ref{sec:using}).}
\end{itemize}

All of the example data used in this paper, along with links to EFSM inference tool can be found online.\footnote{\label{url}\url{http://www.cs.le.ac.uk/people/tg75/efsmdata/}}. 

\section{Background and Related Work}
This section discusses the problem that ITPs demand a significant amount of time and effort to complete the proof process. After reviewing some previous work focussed on aiding with this problem, we introduce Extended Finite State Machines as a possible mechanism to improve proof development by representing existing proofs by means of a descriptive, sequential model.
\label{sec:background}
\subsection{Interactive Theorem Provers}
The expressiveness of interactive theorem provers has led to an abundance of formal proofs becoming available in proof libraries that are distributed with each system. These proof libraries can then be used during the development of new proofs. As with conventional programming languages, developers can build up and exchange their own libraries of proofs to suit their particular domain. Nevertheless, most non-trivial proofs still require an extensive manual effort - Wiedijk states that it takes as long as one week to formalize one page from an undergraduate mathematics textbook \cite{Freek08}. 

The emergence of tools such as Sledgehammer \cite{Sledgehammer07} has reduced this effort to an extent, by enabling Isabelle to call on powerful external Automated Theorem Provers (ATPs) that attempt to solve the goal automatically. Although such tools have been proven to have great value, they require extensive research into the translation between different logics as ATPs utilize different logics to the higher order varieties typically used in ITPs \cite{Meng08}. An empirical study of Sledgehammer \cite{Bohme10} indicated a success rate of 45\% at proving goals from 7 Isabelle theories (known collectively as the Judgement Day benchmark).

Another method of reducing human intervention is called \emph{proof planning} \cite{Bundy88}. Proof planning allows the encoding of reusable strategies that are used to guide proof search - for example many inductive proofs follow a similar pattern which can be encoded as a proof plan. A typical proof plan contains preconditions to state when then plan is applicable, a postcondition stating the effects of executing the proof plan, and the relevant proof steps to apply. Proof planning has been implemented for Isabelle by a tool called IsaPlanner \cite{Dixon03}.

An interesting strand of research is the use of machine learning techniques to improve theorem proving by guiding the proof search or suggesting hints to users. One area that has benefitted greatly from machine learning is the \emph{premise selection problem} \cite{Alama11,Flyspeck12}. Informally, this is the problem of selecting useful premises (from a large collection) to automatically solve a new proposition. By utilizing machine learning techniques, the performance of ATPs on large theory reasoning significantly improved on the state of the art \cite{Kaliszyk13}. Recently, machine learning capabilities have been added to Sledgehammer \cite{Kuhlwein13}. By using the same empirical study (Judgement Day \cite{Bohme10}) that was used to evaluate the original Sledgehammer, it was shown that using machine learning can improve the percentage of completed proofs to 70\%.

Recently, there has been the emergence of a tool called ML4PG \cite{ML4PG13} that uses statistical machine learning techniques to identify commonalities between Coq proofs. Given a proposition that a user is trying to prove, ML4PG can automatically identify clusters of existing lemmas that follow a common proof strategy. The user can then interpret the results and formulate the proof themselves by analogy, using the suggestions provided. ML4PG has been shown to work in a variety of areas such as computer algebra \cite{ML4PGAlg} and industrial proofs \cite{ML4PGACL2}.

\subsection{Motivating Scenario}
To motivate our work, let us consider the following scenario. A novice user is trying to prove (in this case using Coq) the following \emph{app\_nil\_l} proposition stating that an empty list appended to a list {\tt l} should result in {\tt l}:

\vspace{2mm}
\noindent\texttt{Lemma app\_nil\_l: forall l:list A, [] ++ l = l.}

\vspace{2mm}
We assume that the user will be aware of the possible proof methods available in Coq. However it may be unclear how one would sequence these proof methods to arrive at a successful proof. One approach that a user could try might be to manually scour existing proofs to find a sequence of proof steps that will prove \emph{app\_nil\_l}. However, keeping track of the relevant proofs and identifying useful reasoning patterns is time-consuming.

Even for this relatively simple example, finding a proof requires some careful manual processing of the relevant proof libraries. In large scale developments, the task of manually searching through proof corpora to identify the correct steps is generally not practical. We present an automated method based on state machine inference techniques. By providing examples of successful proofs we can generate a model capturing all of the reasoning patterns that occur within the chosen corpora of proofs. This model can then be used to drive the proof search by presenting options to the user about which proof steps to try.

\subsection{State Machine Inference}
State machine inference techniques can address the challenge of identifying the rules that govern a particular sequencing of events. The problem of deriving a model from sequences of events was introduced by Gold in 1967 \cite{Gold67}. Since then it has become a well-established problem, spawning several families of algorithms for different types of models, learning settings and problem domains. The archetypal model for sequences of events is the Finite State Machine (FSM).

\begin{definition}\textbf{\emph{Finite State Machine}}
\label{def:fsm}
 A Finite State Machine (FSM) is defined as a tuple $(S,s_0,F,L,T)$. $S$ is a set of states, $s_0 \in S$ is the initial state, and $F \subseteq S$ is the set of final states. $L$ is as defined as the set of labels. $T$ is the set of transitions, where each transition takes the form $(a,l,b)$ where $a,b \in S$ and $l \in L$. When referring to FSMs, this paper assumes that they are deterministic.
\end{definition}

In the past 40 years numerous algorithms have been developed to infer FSMs (equivalently regular grammars) from observed sequences of events \cite{Biermann72,Lang98,WalkinshawStam13}. These sequences are referred to as \emph{traces}, and are recorded from the system under analysis. The challenge is to derive from the set of traces a FSM that accurately captures the set of all valid sequences of events, even if they do not belong to the initial set of traces.

Such techniques have previously been applied to proof planning. Jamnik \emph{et al.} \cite{Jamnik03} used an Inductive Logic Programming technique to infer what are ultimately regular expressions from well chosen sets of proof methods. For example, if we have the following two proofs (where \texttt{a-d} are proof methods): \texttt{[a, a, c, d]} and \texttt{[a, b, d]} they may be generalized as the following: \texttt{[a*, [b|c], d]}.  

The value of even such a basic model is intuitive. Jamnik \emph{et al.} demonstrated that the models were useful for the automatic generation of new proofs in the $\Omega$\textsc{mega} prover. However, the proof steps that were learned in the examples do not contain any parameters, they are simply method names meaning that this kind of model is too basic to be applied to provers such as Isabelle and Coq. A proof in either of these provers not only relies on the sequencing of the proof steps, but also the values of the parameters provided to these steps.

To combat this problem, this paper explores the use of the Extended Finite State Machines \cite{Cheng93} as a means of modelling examples of successful proofs. EFSMs extend the traditional FSM. Transitions are labelled with guards on an underlying data store (although the update functions on the store are not explicitly modelled).

\begin{definition}\textbf{\emph{Extended Finite State Machine}}
\label{def:efsm}
 An Extended Finite State Machine (EFSM) $M$ is a tuple $(S,s_0,F,L,V,\Delta,T)$, where $S,s_0,F$ and $L$ are defined as in a conventional FSM. $V$ is a store represented by a set of variables, and $v$ represents a set of variable values. $\Delta$ is the set of \emph{data guards}, where each guard $\delta$ takes the form $(l,v)$, where $l \in L$, $v \in V$ is the set of possible data variable configurations specified by the guard. The set of transitions $T$ is an extension of the conventional FSM version, where transitions take the form $(a,l,\delta,b)$, where $a,b \in S$, $l \in L$, and $\delta \in \Delta$. 
\end{definition}
\begin{definition}\textbf{\emph{Traces}}
\label{def:traces}
A \emph{trace} $T=\langle e_0,\ldots,e_n\rangle$ is a sequence of $n$ trace elements. Each element $e$ maps to a tuple $(l,v)$, where $l$ is a label representing the names of function calls or input / output events, and $v$ is a set of corresponding variable values (this may be empty).
\end{definition}

In recent years, algorithms have been developed to infer EFSMs from traces of events \cite{Lorenzoli08,Walkinshaw13}, where events are paired with a selection of variable values. In this work we choose the EFSMInfer tool by Walkinshaw \emph{et al.} \cite{Walkinshaw13}, which has been shown to be reasonably accurate when applied to the task of reverse-engineering models of software modules. We provide a brief overview of the essential steps of the approach below.

Given a set of traces (see Definition \ref{def:traces}), the approach first infers the guard conditions. For each symbol $l \in L$ the trace is scanned, identifying every instance where $l$ is applied, the variable values $v$ at that instance, and the label of the subsequent step in the trace. This is used to construct a training set where, with the use of standard machine learning algorithms (e.g. decision tree learners like \cite{Quinlan93}), it is possible to construct a model that predicts from a given pair label and data configuration what the subsequent label will be. In terms of EFSMs, this gives us $L,V,\Delta$, and implies some constraints on the order in which particular configurations of labels and variables can occur.

The subsequent task is to derive an underlying state transition model that obeys and incorporates these data guards. To achieve this EFSMInfer applies an augmented version of the standard FSM state merging algorithm (Lang's Blue-Fringe algorithm \cite{Lang98}). The set of traces is first arranged as a prefix tree \cite{WalkinshawStam13}, where traces with the same prefix also share the same path from the root. Subsequently, states in the tree are merged according to the likelihood that they represent the same state, based on the similarity of their outgoing paths.

Since this model incorporates data, the merging process includes a step to ensure that the model remains consistent with the data guards. Each transition in the tree is mapped to its corresponding variable configurations. Pairs of states are only merged if the resulting model completely obeys the data classifiers (guard conditions) that were obtained in the previous step. If the inferred data model predicts that the data value for a given transition is followed by a label $l$, any merge involving the target state can only occur if the resulting state machine contains an outgoing transition that is labelled by $l$. After each merge, the resulting state machine is further post-processed to ensure that each transition is deterministic \cite{Walkinshaw13}. 

EFSMInfer has several optional parameters. The most important parameter is the choice of data classifier algorithm, which is used to infer the guards on the transitions. For this, EFSMInfer incorporates several standard algorithms that were implemented as part of the Weka \cite{Weka09} toolset. In our experiments, we will adopt the default parameters in EFSMInfer.

\section{Inferring EFSMs from Proof Corpora}
\label{sec:efsm}
This paper shows how the EFSMInfer tool can be used to derive models from proofs that not only describe the possible sequences of proof steps that have been used in existing proofs, but also the necessary parameter values associated with these proof steps. Although previous work on EFSMs has focussed on program execution traces, they also appear to be well suited to the domain of interactive proofs where we want to capture the interplay between control (proof steps) and data (parameters).

In this section, we describe the process of inferring EFSMs from proofs, and provide a description about how such a model can be interpreted. We begin by showing how existing proofs can be converted into traces, before demonstrating how the model is inferred from these proof traces. The example model shown in this section is for a set of proofs called \texttt{ListNat}, that contains proofs about simple properties of lists and natural numbers.

\subsection{Turning existing proofs into proof traces}
A typical tactical proof script\footnote{Although this work concentrates on Coq, the method in principle can be applied to other ITPs.} contains many examples of propositions that have been proven, along with the sequence of proof steps that the expert user entered to complete the proof. Each proof step has the structure: $proof\_method\;p_{1}\ldots p_{n}$ where $proof\_method$ refers to a Coq command (e.g. \texttt{rewrite, apply, intros}) and $p_{1}\ldots p_{n}$ constitutes the parameters provided to the Coq command. The parameters refer to many different entities such as existing lemmas, rewrite rules or may be related to variables in the goal. 

As shown in Table \ref{table:traces}, the encoding of Coq proofs is a straightforward translation into the trace format shown in Definition \ref{def:traces}. With respect to the tuple of labels and variables $(l,v)$, the $proof\_method$ would correspond to $l$ whilst the parameters $p_{1}\ldots p_{n}$ correspond to $v$. If a proof method doesn't have any parameters provided to it, we indicate this by appending 0 to the end of the proof method (i.e. in Table \ref{table:traces} we see $intros_0$). Also, if proof steps are part of a combination, which is denoted by the presence of a semicolon separating individual proof steps, we encode this information as part of the trace. If two proof steps are put in combination, it means that the first proof step is applied, and then the next one applied to \emph{all} subgoals generated. Including this information in the model is useful so that we know when applying proof steps whether they should be combined.

\begin{table}
\centering
\caption{Original proof and proof trace for an example lemma}
\begin{tabular}{ c  c }
(a) Proof Script & (b) Trace\\
&\\
\begin{minipage}{0.45\textwidth}
\begin{Verbatim}[fontsize=\scriptsize]
Lemma ex : (n*m = O)->(n=O)\/(m=O).
  intros.
  induction n.
  tauto.
  simpl in H.
  right.
  assert (m <= O);
  try omega.
  rewrite <- H.
  auto with arith.
Qed.
\end{Verbatim}
\end{minipage} & \begin{minipage}{0.55\textwidth}
\centering
\begin{tabular}{l|l|l}
Event $e$ & Label $l$ & Values $v$\\
\hline
$e_0$ &intros$_0$& $\langle \rangle$\\
$e_1$ &induction & $\langle p_1 = ``n" \rangle$\\
$e_2$ &tauto$_0$ & $\langle \rangle$\\
$e_3$ &simpl & $\langle p_1 = ``in\;H" \rangle$\\
$e_4$ &right$_0$ & $\langle \rangle$\\
$e_5$ &assert & $\langle p_1 = ``m\leq 0", p_2 =``;" \rangle$\\
$e_6$ &try & $\langle p_1 = ``omega" \rangle$\\
$e_7$ &rewrite & $\langle p_1 = `` \leftarrow H" \rangle$\\
$e_8$ &auto & $\langle p_1 = ``with\;arith" \rangle$
\end{tabular}
\end{minipage} \\
\end{tabular}
\label{table:traces}
\end{table}

\subsection{Inferring the model}
After converting each proof into its corresponding trace, it becomes possible to infer a model from a collection of these traces. We choose the standard configuration for the EFSMInfer tool and, for the sake of illustration select the J48 decision tree learner (a Weka implementation of the C4.5 algorithm \cite{Quinlan93}). Having chosen the classifier we can run the EFSMInfer tool and generate an EFSM.

\begin{figure}
\centering
\begin{minipage}[b]{0.2\textwidth}
\begin{verbatim}
   MODEL FOR:induction
   J48 pruned tree
   ------------------
   (p1 = n): simpl 
   (p1 = a): simpl 
   (p1 = m): trivial 
   (p1 = l): simpl 
\end{verbatim}
\vspace{2cm}
\end{minipage}
\hspace{2cm}\begin{minipage}[b]{0.5\textwidth}\includegraphics[scale=0.14]{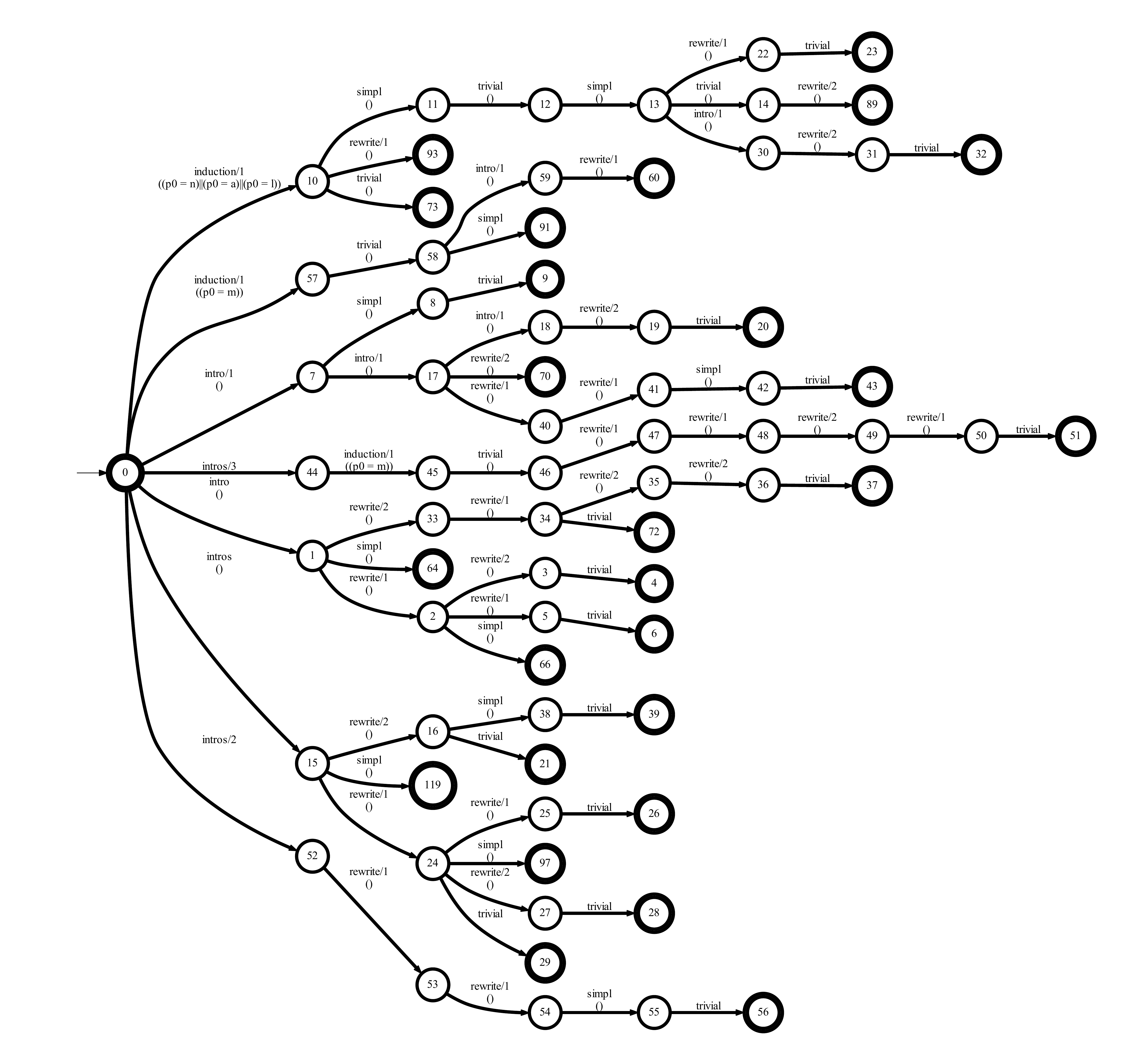}\end{minipage}\\
(a) Data rules for {\tt induction} \hspace{4cm} (b) Prefix tree

\hspace*{-1cm}\includegraphics[scale=0.101]{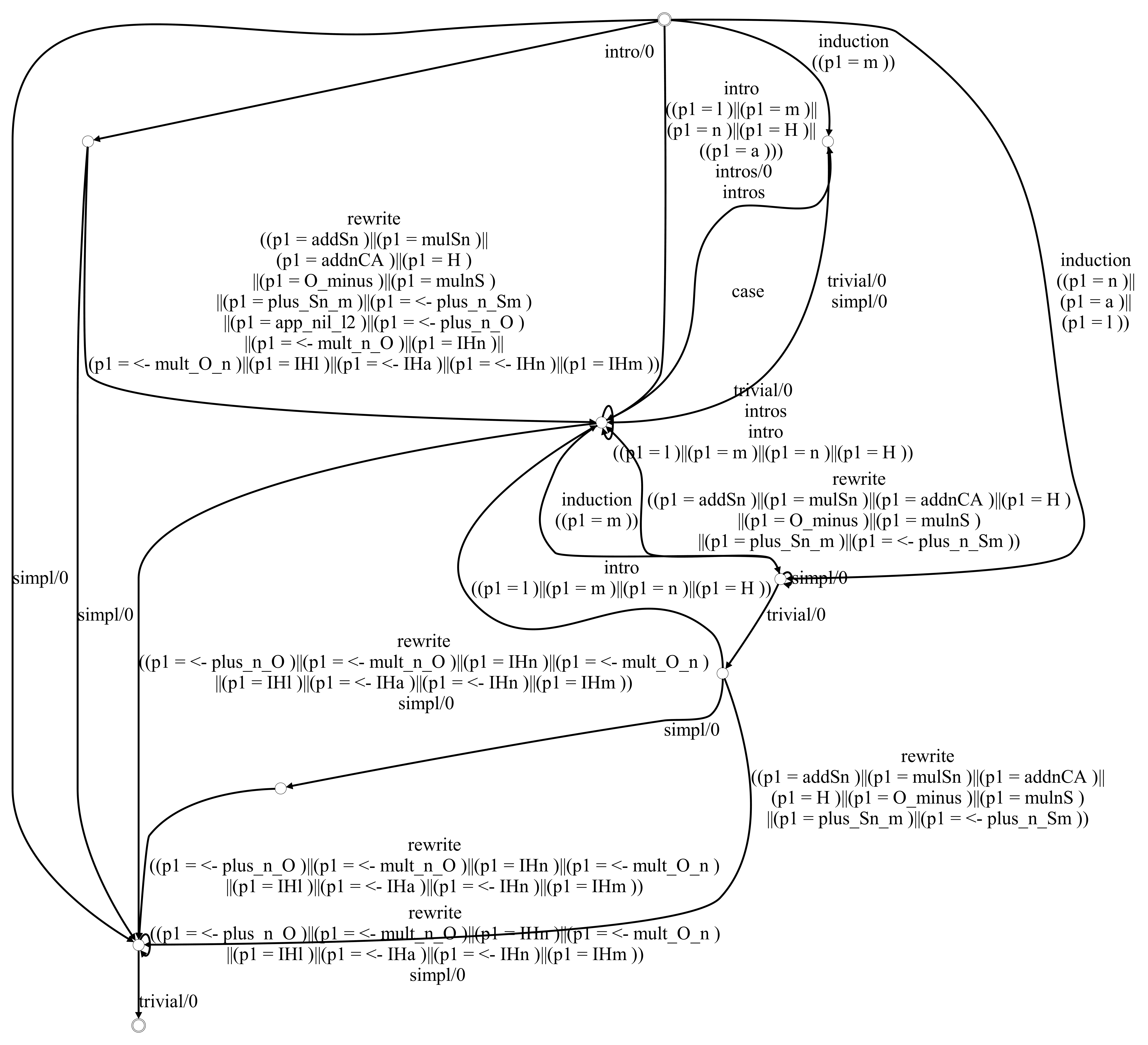}\\
(c) Inferred EFSM
\caption{PTA and inferred EFSM for {\tt ListNat} traces. }
\label{fig:model}
\end{figure}

To begin with data classifiers are inferred that, for each $proof\_method$, produce a function that uses the parameters to predict the subsequent transition in the model. An example data classifier can be seen in Figure \ref{fig:model}(a) for the {\tt induction} proof method. The data classifier is interpreted as follows: if the parameter $p_1$ is equal to \emph{n,a} or \emph{l}, then the subsequent proof method should be {\tt simpl}. If $p_1$ is equal to \emph{m} then the following proof method should be {\tt trivial}. Although not the case here, the C4.5 algorithm can produce more complex trees of if-then-else rules governing the possible value ranges for parameters if necessary.

Once the data classifiers have been inferred, the state merging can commence. Initially, the set of proof traces is arranged as a prefix tree. The tree for our example is shown in Figure \ref{fig:model}(b). The labels are unreadable, but the purpose is merely to give an intuition of what the tree might look like, and to illustrate the ensuing state merging challenge. Each transition in the tree is associated with a label (which is linked to one of the inferred data classifiers), along with the variable values that correspond to that transition. The inference challenge for the merging algorithm is to select compatible pairs of states to be merged. These states should have similar outgoing paths, should not entail the merging of states that are incompatible (e.g. accepting and rejecting), and should not raise contradictions with the inferred data classifiers (as discussed in Section \ref{sec:background}).

The final EFSM is shown in Figure \ref{fig:model}(c). The constraints on the transitions detail the parameter configurations that are associated with each transition. The model is deterministic; for any state there is never more than one outgoing transition for a given combination of label and variable configuration.

\section{Using EFSMs in Interactive Theorem Proving}
\label{sec:using}
This section seeks to determine the potential value of inferred EFSMs as a guidance mechanism for users of interactive theorem provers. To assess the EFSMs when applied to proofs, we conduct an automated experiment to measure the accuracy of the models, whilst also producing a more informal, qualitative case study to show how the models can be used to manually derive new proofs. Our notion of accuracy revolves around the inferred EFSM's ability to distinguish whether a proof should be accepted by the model or not. We conclude the section by highlighting some future work to improve our current technique.

\subsection{Assessing the accuracy of inferred EFSMs}
\paragraph{Measuring accuracy.}
Measuring the accuracy of inferred models is challenging, especially in the absence of ``gold-standard" models that could be used as a reference. In machine learning this problem is common. One of the most popular evaluation techniques that can be used in such a situation is known as \emph{k-folds cross validation} \cite{Kohavi95}. The dataset is randomly partitioned into $k$ non-overlapping sets (also known as folds). Over $k$ iterations all bar one of the folds are used to infer a model, and the remaining fold is used to evaluate the model according to some metric (discussed below). For each iteration a different fold is used for the evaluation. The final accuracy score is the average of the $k$ accuracy scores.

Of course, given the probability that the given set of proofs is not ``rich enough", the accuracy score cannot be interpreted as an \emph{absolute} score of the accuracy of the model. However, if we accept that the test set captures a representative set of proofs for a given domain, then the resulting scores can be interpreted as being at least indicative of the actual accuracy score.

To assess `accuracy' there are many metrics that we can choose as a measurement such as precision, sensitivity and specificity. All are computed from the set of true-positives (TP), true-negatives (TN), false-positives (FP) and false-negatives (FN). In this experiment we choose sensitivity (TP/(TP+FN)), and specificity (TN/(TN+FP)).

\paragraph{Negative examples.} As indicated above, fully assessing sensitivity and specificity implies the existence of ``negative" proof traces - traces that do not correspond to valid proofs (and therefore should not lead to accepting states in an inferred model). For the purposes of this evaluation we have selected some of the positive examples and mutated the sequences of proof steps by randomizing them. In addition to this, we provide sequences of proof steps from theories that are different from the ones we have inferred a model from. In practice, these negative examples could be captured from proof attempts that have failed to prove a proposition. In each experiment, we use approximately 30 negative examples.

\paragraph{Evaluation process.} To get an idea how accurate our inferred models are we use $k$-folds cross validation each of our datasets. We set the number of folds $k=5$ to ensure that we have an adequate sized evaluation set at each iteration. At each iteration of the $k$-folds an EFSM is inferred from the traces contained in the training set. We then run the traces from the evaluation set and the negative traces through the model, logging whether they are accepted or not. From this we can compute sensitivity and specificity. Since the inference of the guards in the EFSM depends on the selection of a suitable data classifier algorithm, we repeat the experiment for five data classifiers provided in EFSMInfer.

\begin{table}[t]
\centering
\caption{Proof libraries, and the accuracy of the inferred models.}
\begin{tabular}{lll|cc}
\hline
Data Set & Proofs & Lines & Sensitivity & Specificity \\
\hline
ListNat & 70 & 660 & 0.84  & 0.81 \\
Bool & 100 & 809 & 0.95  & 0.55 \\
Coqlib & 100 & 1326 & 0.22  & 0.96 \\
Values & 85 & 1188 & 0.24  & 0.98 \\
\hline
\end{tabular}
\label{tab:datasets}
\end{table}

\paragraph{Data sets.} We chose four sets of Coq proofs, which are listed (along with the number of proofs and lines of code) in Table \ref{tab:datasets}. \texttt{ListNat} contains proofs regarding the basic properties of lists and natural numbers. \texttt{Bool} contains proofs about boolean values. To complement these datasets, we also chose two theories contained in CompCert \cite{Leroy09}, which is a formally verified C compiler. The \texttt{Coqlib} theory contains proofs about functions used throughout CompCert, whilst \texttt{Values} focuses on proofs related to run-time values. All of our datasets are composed of hand curated proofs so that the models don't simply contain calls to automated tactics that may solve the goal instantly.

\subsubsection{Results.}
For each proof set, the choice of data classifier algorithm made a negligible difference to the results. The five classifiers (all part of the Weka distribution \cite{Weka09}) were J48, NaiveBayes, NNGE, AdaBoostDiscrete and JRIP. Our results in Table \ref{tab:datasets} show the values obtained from using the J48 classifier. For all systems apart from \texttt{Bool}, the specificity measures are all 85\% and above. In these cases there were very few false-positives (meaning that a low proportion of negative examples were falsely accepted by the model).

The sensitivity values vary substantially depending on the dataset used. The \texttt{ListNat} and \texttt{Bool} datasets have reasonably high sensitivity values (both over 80\%), indicating that they were good at predicting new proofs that did not belong to the training data. \texttt{Coqlib} and \texttt{Values} had low sensitivity scores, meaning that the inferred models failed to predict a large proportion of proofs that were not in the training set. 

In the cases of \texttt{Coqlib} and \texttt{Values}, the low sensitivity scores are not particularly surprising and is to an extent inevitable. Whereas the proofs in \texttt{ListNat} and \texttt{Bool} are relatively homogeneous because they are concerned with specific, simple data structures, the proofs in \texttt{Coqlib} and \texttt{Values} are highly diverse and have less common reasoning patterns than the other libraries. \texttt{Coqlib} provides a general library of proofs that are intended for use in almost \emph{any} context. \texttt{Values} provides proofs that apply to the values of variables in a compiler and, given that CompCert is entirely concerned with a compiler verification, plays a central role in the diverse range of contexts within CompCert.

In such cases, the EFSMInfer tool is inevitably only provided with a small fraction of the proofs that are required to constitute a truly `representative' training set. Accordingly, the tool is bound to under-generalize, resulting in models that are too conservative; the proofs that they predict are largely valid, but they invariably miss out many other proofs that are in fact valid.

\subsection{Case Studies}
Although the results from the previous section provide us with a qualitative assessment of the accuracy of the inferred models, they only provide a limited insight into the practical value of the models from a user's perspective. We conclude this section with a detailed walk through the process of how a user can derive a proof using an EFSM as guidance. We show two case studies that demonstrate the process of using an EFSM during the proof process.

In the subsequent examples, we model the following scenario. Let us assume that we have a collection of existing proofs available; \texttt{ListNat} contains proofs about lists and natural numbers and is used in case study 1. The \texttt{Bool} dataset contains proofs about boolean values, which we use in case study 2. We then suppose that we task a Coq user to prove one of the lemmas in the dataset (and allow them to use the remaining proofs to infer a model from). We demonstrate that in each case we can be led to a proof using the model as guidance. We then compare the EFSM based proof with the original proof from each dataset. In both cases, we see interesting results when we compare.

\subsubsection{Example 1}
Let us refer back to the motivating example from Section \ref{sec:background}, where the user is tasked with proving the \emph{app\_nil\_l} proposition, which is part of the \texttt{ListNat} dataset. As an exercise, let us assume that the user has been given the remainder of the \texttt{ListNat} proofs (minus the proof for \emph{app\_nil\_l}). Figure \ref{fig:model}(c) shows the EFSM associated with this example and was inferred from every proof in \texttt{ListNat} minus \emph{app\_nil\_l}. The process that one might follow to derive a proof from the model is as follows: 
\begin{itemize}
\item{Our main choices to start the proof are \texttt{induction} or \texttt{intros}. We know that typically proofs containing lists begin with induction, and the model also suggests parameter $p_1=$\texttt{ l}, so we select the first step of the proof as \texttt{induction l}.}
\item{The first subgoal that needs proving is the base case showing that appending 2 empty lists together results in the empty list. The options that the model suggests are the following - \texttt{trivial}, \texttt{simpl} or \texttt{rewrite}. This particular subgoal is a simple equality, so it suffices to choose \texttt{trivial} as the next proof step.}
\item{We can now move on to the inductive step. The model then presents us with 3 more options - \texttt{intro}, \texttt{simpl} or \texttt{rewrite}. None of the parameters suggested for \texttt{rewrite} seem to be applicable, they are more suited to natural number proofs. There is nothing we can introduce, so we choose to simplify using \texttt{simpl}.}
\item{There is only one possible step that can follow, which is \texttt{rewrite}. Besides a couple of existing lemmas regarding natural numbers, the model seems to be suggesting rewriting the inductive hypothesis. By analogy with the model we choose the parameter $p_1=$\texttt{ <- IHl}.}
\item{Finally, we can complete the proof (and arrive at an accept state) by using \texttt{trivial}.}
\end{itemize}

We have shown by using an EFSM that one way of solving \emph{app\_nil\_l} would be to use the following sequence of proof steps: 

\vspace{2mm}
\noindent\texttt{induction l. trivial. simpl. rewrite <- IHl. trivial.} 
\vspace{2mm}

\noindent So how does this proof compare against the original proof for the same proposition in ListNat? The existing proof was the following:

\vspace{2mm}
\noindent\texttt{intro l. case l. simpl. trivial. intros a0 l0. simpl. trivial.} 
\vspace{2mm}

\noindent Interestingly, the proof found by using the EFSM was two steps shorter, and also required less effort in identifying the parameters required for the proof steps. Additionally, the sequence found from traversing the EFSM was (at least not in its entirety) part of the training data, and was only found as a result of inferring an EFSM.

\subsubsection{Example 2} In our second example, we try to prove the following proposition: \texttt{negb(b1 $||$ b2) = negb b1 \&\& negb b2}, which states that (for two boolean values b1 and b2) if b1 or b2 is false, then b1 is false and b2 is false. We infer a model from all of the other proofs available in the \texttt{Bool} dataset. The corresponding model for this example can be found on the authors webpage\footnotemark[1]. The process of using the model to arrive at a proof is the following:
\begin{itemize}
\item{To begin the proof, the model suggests either \texttt{destruct} or \texttt{intros}. We try the \texttt{intros} path first as there are quantified variables that we can introduce, but we are then led to a state where nothing is applicable. So we must use \texttt{destruct} instead. There are numerous parameters that are suggested, but we see that we have \texttt{b1} and \texttt{b2} in our goal, so it makes sense to choose parameters that include one of these. We decide to set $p_1$ = \texttt{b1} and $p_2$ =\texttt{ ;} to make the proof step \texttt{destruct b1 ;}}
\item{We are presented with a number of options, most of which we can rule out due to not being applicable e.g. \texttt{rewrite, case}. We do have a boolean \texttt{b2} in our goal, so we follow the suggested step - \texttt{destruct b2 ;} }
\item{At the next node, there are 2 possible paths. One involves the \texttt{rewrite,case} steps that we still cannot apply. We take the path that uses the simplifying method \texttt{simpl} and the suggested parameters which are $p_1$ ={\tt in |- *} and $p_2 =$ {\tt ;}}
\item{Finally, the model suggests {\tt trivial} or {\tt reflexivity} to complete the proof. Either of these lead to the proof, but we choose the {\tt trivial} method for the purposes of this example.}
\end{itemize}
We have again been led to a proof by following the guidance provided by the EFSM. The proof that corresponds to the sequence above is the following:

\vspace{2mm}
\noindent\texttt{destruct b1; destruct b2; simpl in |- *; trivial.}
\vspace{2mm}

\noindent The original proof from the Bool dataset corresponds to the following proof steps:

\vspace{2mm}
\noindent\texttt{intros; destruct\_all bool; simpl in |- *;trivial;try discriminate}
\vspace{2mm}

\noindent Although only 1 step shorter this time, we have again shown that using an EFSM to complement the proving process can yield useful results. In this particular example, we have shown that there is a smaller number of distinct proof methods used in the newly found EFSM based proof than in the original one.

\subsection{Threats to validity}
It is important to bear in mind that these results are primarily intended to be indicative, and as such there are several elements in the design that could potentially invalidate the results. Firstly, we have only chosen four data sets for our experiments. Clearly there may be other collections of proofs that could lead to much better or worse results than the ones described here. Nevertheless, we chose these data sets to ensure a highly diverse selection that covered a wide broad variety of examples. Another potential threat is the generation of the negative examples that factor into the calculation of sensitivity and specificity. By manually inspecting the generated examples we tried to not select negative traces that were too easy to identify as such.

\subsection{Improvements and Future Work}
Although we have shown that inferring models can be useful in the proof process, we haven't yet discussed the current limitations of the approach. We have identified the following areas where our EFSM-based approach can be improved, and in doing this can lead us towards our overall aim, which is to automatically complete proofs using EFSM-based approaches. 

The way we choose to represent parameters in the EFSMs can be improved. Currently everything is treated entirely textually, so an interesting avenue for future work would be to abstract away from the actual variable names and investigate the inclusion of the \emph{types} of the variables instead. This would help to simplify the models, whilst also making them applicable to a larger range of propositions.

Another limitation is being able to identify the relevant paths through the model for any given proof. From a user's perspective, when presented with a small model such as the one shown in Section \ref{sec:efsm}) they can simply evaluate the options and each step and make an informed choice. We are ultimately interested in a system that can execute the EFSM automatically to derive proofs. This could be done in a number of ways, for example by using a Breadth-First Search of the EFSM to check the applicability of proof steps, or by using evolutionary algorithms.

The negative information that we used in the experiments is not entirely accurate, in the sense that a more robust selection of negative examples could be actual failed proof attempts. In addition to improving the quality of the negative examples, we are also interested in the incorporation of this negative information within the model \cite{Walkinshaw09}. By including this information within the model, we may be able to infer much more accurate models of proofs.

A final consideration is the selection of proofs that we infer EFSMs from. The approach we use in this paper is to select \emph{similar} proofs in the sense that proofs are grouped together because they all deal with a similar data structure, or are contained within the same theory file. An interesting addition to our tool would be to make use of proof filtering tools such as ML4PG \cite{ML4PG13}. By using ML4PG, we could inspect the proof obligation that we are trying to prove, before being presented with the most relevant proofs (as suggested by ML4PG). We can then use these suggestions as input to EFSMInfer, instead of a collection of manually selected proofs.

\section{Conclusion}
\label{sec:conclusion}
We have shown how EFSMs can be derived from existing proof corpora. These state machines have proven to be useful as they can reduce large, complex proof files into a more manageable, concise representation. In our evaluation, we have demonstrated that the models are reasonably accurate and that they can be used to derive new proofs. We have also shown that in comparison to existing proofs, the EFSM based ones can be shorter and less complex that the original. The models not only show a user the possible sequencing of proof methods (which is valuable enough information on its own), but also help to suggest the parameters that may be useful in completing a proof. Finally, we have highlighted some areas for improving our technique in the future.
\bibliographystyle{splncs03}
\bibliography{cicm-bib}

\end{document}